\documentclass{elsart}
\input epsfig.sty
 \begin{document}
 \runauthor{Mukherjee \& Chiang}

 \begin{frontmatter}
 \title{EGRET Gamma-Ray Blazars: Luminosity Function and Contribution to the 
 Extragalactic Gamma-Ray Background}

 \author[NY]{R. Mukherjee}
 \author[JILA]{\& J. Chiang }
 \address[NY]{Barnard College \& Columbia University, Dept. of Physics \& 
 Astronomy, New~York, NY 10027}
 \address[JILA]{JILA, University of Colorado, Boulder CO 80309-0440}

 \begin{abstract}
 We describe the properties of the blazars detected by EGRET and summarize the 
 results on the calculations of the evolution and luminosity
 function of these sources. 
 Of the large number of possible origins of extragalactic diffuse $\gamma$-ray 
 emission, it has been postulated that active galaxies might be one of the 
 most likely candidates. However, some of our recent analysis indicate that 
 only 25\% of the diffuse extragalactic emission measured by 
 SAS-2 and EGRET can be attributed to unresolved $\gamma$-ray blazars. 
Therefore, 
 other sources of diffuse extragalactic $\gamma$-ray emission must exist. 
 We present a summary of these results in this article. 
 \end{abstract}

 \end{frontmatter}

 \section{Introduction}
 EGRET has detected a total of 66 active galactic nuclei (AGN) 
 in high energy ($> 100$ MeV) gamma rays since the launch of CGRO in April 
 1991 (Hartman et al. 1999). These sources all appear to be members of the 
 blazar class of AGN (BL Lac objects, highly polarized ($ >$ 3\%) quasars 
 (HPQ), and optically violently variable (OVV) quasars) and are radio-loud 
 sources with flat-spectrum at radio bands. Many of the blazars exhibit 
 variability in their $\gamma$-ray flux on timescales of several days to months 
 (McLaughlin et al. 1996, Mukherjee et al. 1997). The photon spectra of the 
 blazars in the energy range 30 MeV to 30 GeV are generally well 
 represented by power laws in energy with photon spectral indices in the 
 range 1.4 to 3.0. The sources have non-thermal continuum 
 spectra with the $\gamma$-ray luminosity exceeding those at other frequencies in 
 most cases. 
 The high $\gamma$-ray luminosities of the 
blazars suggest that the emission is 
 likely to be beamed and, therefore, Doppler-boosted along the line of sight. 
 The spectral energy distribution of blazars can be modeled 
 as follows: the radio to UV emission can be explained as synchrotron 
 emission from relativistic electrons in a uniform relativistically moving 
 plasma. The high energy emission is due to the inverse Compton scattering of 
 seed photons off the relativistic electrons, although the source of the soft 
 photons still remains unresolved (see Hartman et al. 1997 for a review). 

 In this article we summarize the luminosity function and evolution properties 
 of the EGRET blazars and use the results to examine the contribution of the 
 $\gamma$-ray-loud AGN to the diffuse extragalactic background. 

\section{Luminosity function of EGRET blazars}

The evolution and luminosity function of the EGRET blazars was calculated 
by Chiang \& Mukherjee (1998) using data from the Phase 1 through Cycle 4 
CGRO observations. 
Inclusion in the 1 Jy catalog of K\"uhr et al. (1981) of the 
EGRET blazars was used to account for possible biases introduced by missing 
optical identifications. A 
$V/V_{\rm max}$ test was used to show evidence of evolution. 
Here $V$ is the minimum volume that contains an object with redshift $z$; 
$V_{\rm max}$ is the largest volume that could contain an object with the same 
luminosity, and still be detected at the given flux limit. 
For a limiting significance of detection of $4\sigma$, a value of 
$\langle V/V_{\rm max}\rangle = 0.7$ was obtained, which 
means that we are preferentially detecting more sources at larger redshifts. 
No evidence of a density evolution of EGRET blazars was found.
The evolution is consistent with pure luminosity evolution. 
(That is, the luminosity of the object is changing with time (i.e. redshift), 
while the co-moving number density remains the same). 

The luminosity of a given object as a function of redshift $z$ 
can be described by $L(z) = L_0 f(z)$ 
where $L_0=L(z=0)$. Chiang \& Mukherjee (1998) have discussed several 
different forms for the luminosity evolution function, including the 
power-law and exponential forms. 

The redshift distribution of EGRET blazars was used to characterize the 
low end of the luminosity function better. 
The high end of the luminosity function was fixed by the non-parametric 
estimate mentioned above. 
The redshift distribution of the EGRET data was used to fit both the 
break luminosity and power law index of the low end of the luminosity 
function. 
A likelihood function of the redshift distribution was constructed. 
The probability density for the redshift of a given blazar was computed and 
normalized assuming the flux limit derived for that blazar. 
The data were best fit with a single power law at high 
luminosities and a luminosity cutoff of $1.1\times 10^{46}$ ergs s$^{-1}$. 

Using the lower limit of the de-evolved luminosity function, the
$\gamma$-ray loud AGN contribution to the extragalactic $\gamma$-ray flux
is estimated to be $4.0^{+1.0}_{-0.9}\times 10^{-6}$ photons cm$^{-2}$
s$^{-1}$ sr$^{-1}$. The sky-averaged flux contribution of identified EGRET
blazars is $\simeq 1\times 10^{-6}$ photons cm$^{-2}$ s$^{-1}$ sr$^{-1}$.
Contribution to the diffuse background by unresolved blazars, therefore, is
$\sim 3.0^{+1.0}_{-0.9}\times 10^{-6}$ photons cm$^{-2}$ s$^{-1}$ sr$^{-1}$. The
extragalactic diffuse flux for $E>100$ MeV estimated by Sreekumar et
al. (1998) is $\simeq 1.36\times 10^{-5}$ photons cm$^{-2}$ s$^{-1}$
sr$^{-1}$.  We therefore find that blazars cannot account for all of the
diffuse extragalactic $\gamma$-ray background at the energies considered.

\begin{figure}[t!]
\centerline{\epsfig{file=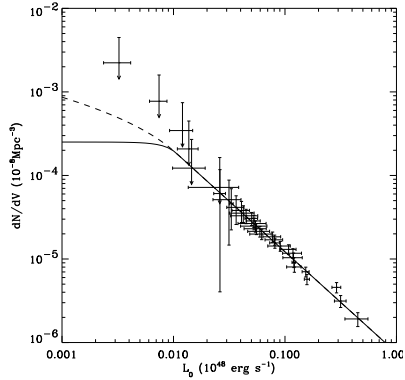,height=2.0in,bbllx=20pt,bblly=-20pt,bburx=420pt,bbury=350pt,clip=.}}
\caption{Cumulative luminosity function derived from the de-evolved 
luminosities. The data points are the luminosity function estimates using the 
smoothed nonparametric method of Caditz \& Petrosian (1993). The solid line 
is the cumulative luminosity function derived from a broken power-law 
differential distribution (Chiang \& Mukherjee 1998). }
\end{figure}

\section{Summary}

The luminosity function and evolution properties of $\gamma$-ray-loud
blazars imply that only $\sim$ 25\% of the diffuse extragalactic emission
measured by SAS-2 and EGRET can be attributed to unresolved $\gamma$-ray
blazars.  This is contrary to other estimates which assume a linear
correlation between the measured radio and $\gamma$-ray fluxes (e.g.,
Stecker \& Salamon 1996).  However, we note that our result is consistent
with recent work by M\"ucke \& Pohl (1998) where the extragalactic diffuse
contribution from blazars is synthesized using a specific blazar emission
model (Dermer \& Schlickeiser 1993) and an extrapolation of the observed
log$N$--log$S$ distribution of EGRET blazars.  As in our study, M\"ucke \&
Pohl make no assumptions regarding supposed correlations between the
$\gamma$-ray fluxes with any other spectral band.  

Our results lead to the exciting conclusion that other sources of diffuse
extragalactic $\gamma$-ray emission must exist.  The spectrum of the
measured extragalactic emission implies that the average quiescent energy
spectra of these sources extend to at least 50 GeV and maybe up to 100
GeV, without a significant change in the slope.  If gamma-ray blazars
continue to make a significant contribution to the diffuse emission at
these energies, then the spectra of the parent relativistic particles in
blazars which produce the gamma-rays must also remain hard to even higher
energies.

\end{document}